# An EEG pre-processing technique for the fast recognition of motor imagery movements

Kalogiannis Gregory
Aristotle University of Thessaloniki
Department of Electrical and
Computer Engineering
Thessaloniki, Greece
gkalogiannis@ece.auth.gr

Kapsimanis George
Aristotle University of Thessaloniki
Department of Electrical and
Computer Engineering
Thessaloniki, Greece
kapsimag@ece.auth.gr

Hassapis George
Aristotle University of Thessaloniki
Department of Electrical and
Computer Engineering
Thessaloniki, Greece
chasapis@eng.auth.gr

*Abstract* — In this paper we propose a new pre-processing technique of Electroencephalography (EEG) signals produced by motor imagery movements. This technique results to an accelerated determination of the imagery movement and the command to carry it out, within the time limits imposed by the requirements of brain-based real-time control of rehabilitation devices, making thus feasible to drive these devices according to patient's will. Based on event related desynchronization and synchronization (ERD/ERS) of motor imagery, the received patient signal is first subjected to the removal of environmental, system and interference noise which correspond to normal human activities such as eye-blinking and cardiac motion. Next, power and energy features of the processed signal are compared with the same features of classified signals from an available database and the class to which the processed signal belongs, is identified. The database classification is done off-line by using the SVM algorithm.

*Keywords* — *rehabilitation, electroencephalography, preprocesss, motor imagery movements*

## I. INTRODUCTION

Injuries and trauma to the human joints, commonly require surgical treatment followed by post-operative physical therapy rehabilitation. Devices, such as continuous passive motion machines (CPM), are used for rehabilitation in hospitals, clinics or general practices and they are important supplement to medical and therapeutic treatment. The goals of rehabilitation are: control post-operative pain, reduce inflammation, joint stiffness, swelling, protect the healing repair or tissue [1] and restore the range of motion in the joint post-operatively. Also such devices contribute to regeneration and blood circulation, prevent thrombosis and embolism phenomena [2].

Their mode of operation is to move injured joint over a range of motion in a circular periodical way defined by the physician [3]. For example in case of elbow and fist joints, these devices impose movement via flexion/extension and/or pronation/supination to the injured joint [4]. Although their indisputable contribution to rehabilitation [5], it is believed that the overall treatment time can be reduced and the overall rehabilitation could be improved if the patient interacts with these devices and their motion is determined according to patient's will. A variety of devices host interesting features such as the functionality to periodically increment the maximum angle on each cycle, the ability to program the unit for precise adjustment of patient-specific therapy values or even to store the programmed therapy parameters into local media. These features allow their easy connection with controllers which can make the devices to follow trajectories determined by processing the generated by the patient brain signals. These signals may include the patient's intensions and will for potential limb movement. A typical structure of a commercial rehabilitation device is illustrated in Fig. 1. The implementation of such an architecture requires fast recognition of the motor imagery movements of the joint in order to create the appropriate control signals. This can be done by processing the EEG data with the purpose of removing the noise and information that is not essential for creating the control signal.

In this paper we demonstrate a new preprocess technique of EEG signals focused on event related desynchronization and synchronization (ERS/ERD) [6] phenomena, fast enough to run within the time limits imposed by the on-line control of the continuous passive devices. It is demonstrated experimentally that a significant computation time reduction is achieved against a typical technique without preprocessing in the recognition of a motor imagery movement.

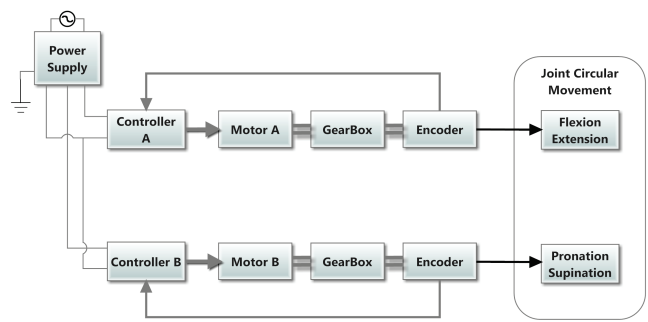

Fig. 1. A typical structure of a commercial CPM device for joint rehabilitation.

In section II the features of EEG signals that are associated with motor imagery movements are explained and the mentioned above proposed technique. This technique involves the recognition of the EEG features related with the motor imagery movement by removing noise and other artifacts which are not related with these movements. In Section III the technique is tested with pre-recorded EEG data and conclusions are drawn which are presented in the same Section. Overall conclusions are presented in Section IV. Section V proposes future work that can be undertaken towards the realization of the complete brain-based control of CPM.

## II. PREPROCESSING OF EEG DATA

Preprocessing of inspected EEG for making easier the extraction of desired features is ponderous. Preprocessing methods used in EEG are very dependent on the goal of the applications. There are some methods that are used very commonly to improve the quality of Signal to Noise Ratio (SNR), such as Common Average Referencing (CAR) [7]. Resampling the data, filtering, bad channel detection, Independent Component Analysis (ICA), epoching continuous data, and epoch rejection are the most common techniques in the preprocessing stage of EEG recordings [8-11].

Since we are interested on motor imagery tasks, we focus on manipulating data that correspond to specific electrodes over the human-patient scalp. These are recordings that are referred to the premotor cortex brain area. Using the typical 10-20 system, the focused electrodes are the $FC_3$, $FC_Z$, $FC_4$, $C_3$, $C_1$, $C_Z$, $C_2$ and $C_4$ as illustrated on Fig. 2.

During imagery motor movements tasks, the so called *mu* and *beta* event-related desynchronization (ERD) and event-related synchronization (ERS) are taking place, allowing us to determine human patient imaginary movement. [12, 13]. These imagery tasks, that can be brief movement imagery or continuous movement imagery, can be recognized as *mu* and *beta* ERD/ERS patterns inside the recorded EEG. Fig. 3, illustrates power time variations of ERD/ERS patterns in EEG signals received from the $FC_3$ electrode, placed on the premotor cortex area of the human brain. Observed power suppression and spikes indicate that the ERD/ERS events of imagery motor movements are taking place. The power spike and suppression sizes and the time window during which they occur, can be used to recognize the type of movement, i.e. open left fist or close right fist, or rotate right arm over the elbow or feet movements.

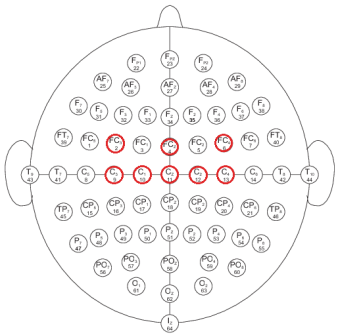

Fig. 2. The 10-20 system with the focused electrodes $FC_3$, $FC_Z$, $FC_4$, $C_3$, $C_1$, $C_Z$, $C_2$ and $C_4$ annotated.

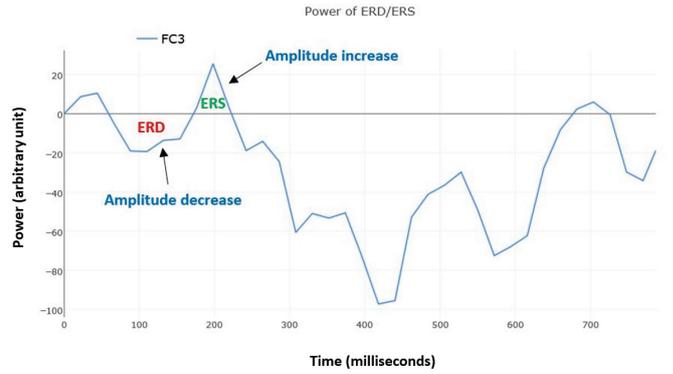

Fig. 3. Event-related desynchronization/synchronization of motor imagery tasks during EEG recording over premotor cortex electrode $FC_3$

### A. EEG data preprocessing focused on ERS/ERD events

During an offline procedure, patient is subjected to imagery motor movement while EEG data is recorded. These data recordings can be used as input data to classify the signals according to their power suppression or spike at the above mentioned time window. Then, once certain signals from a patient are monitored they can be tested to see if they match the power features of a certain signal class, recognizing in this way the specific motor imagery movement of the patient.

However, original recordings of EEG data contain environmental and system noise and interference that corresponds to normal human activity such as eye-blinking and cardiac motion which need to be removed in order to isolate the ERS/ERD events from the rest of the signal. The most obvious type of noise that must be eliminated is the dc Parameter. Therefore, a high pass filter from 0.5 Hz to 90Hz must be applied to the initial data. Furthermore, interferences due to power supply network must be also removed. Therefore, a second band elimination filter at 50 Hz can be used. In order to remove noise that is correlated with eye-movement (EOG) and heart motion (ECG), the blind source separation algorithms (BSS) [14, 15] can be used.

Each recording of offline EEG data, corresponds to a session which may include several individual runs while each run corresponds to a specific task, i.e open/ close left fist. By denoting by $S_w$, the union of all the sessions $S_i$, then $S_w$, is defined as:

$$S_w = \bigcup_{i \in R} S_i \qquad (1)$$

Each run, $R_{nSi}$ is a matrix consisting of *mxn* elements, where m depends on the session selected and n equals to the number of EEG channels $C_n$ that are measured. Each run also is adjusted to a specific task action of the human subject being measured. Since we are focused on imagery motor movements, we are isolating tasks that are correlated to an imagery movement action, for example open or close right/left fist.

Each imagery task $T_i$ has a time of execution $t_{Ti}$. Taking into consideration that we focused on motor cortex events, time of execution $t_{Ti}$ is expanded to include the times of the mu and beta de-synchronization and synchronization phenomena. Also, it is known that in any movement, the ERD event is preceded by the movement-related cortical potential (MRCP) lasting for a few seconds and being associated with delta rhythms (lower than 4Hz.) The sum of all these times comprise a signal epoch of a movement. Therefore, for the purpose of identifying movement, all that is needed is to isolate and extract those sessions and discard the remaining signal.

By denoting by $t_{ERD,Ti}$ the time duration of the ERD case and $t_{ERS,Ti}$ the time duration of ERS, the actual time window of signal observation is:

$$t = t_{ERD,Ti} + t_{Ti} + t_{ERS,Ti} \quad (2)$$

Fig. 4, illustrates the timeline of an event occurred including ERD, ERS and MRCP phenomena.

Still the reduced signal contains interferences that are attributed to eye artifacts. Such noise can be removed by applying the typical Independent Component Analysis (ICA) algorithm [16]. Our final data are a number of vectors, which reflect the Power and Energy of the identified ERD/ERS pattern in the signals. The amplitude of these vectors can be used as a feature to stop or adjust the motion angle of the CPM device. For each $C_n$ EEG channel, we extract two feature vectors that corresponds to power, and energy of the isolated epoch. Power is calculated using the formula

$$P(C_n) = \frac{1}{N}\sum_{n=1}^{N} X(n, C_n)^2 \quad (3)$$

and energy is computed based on the formula

$$E(C_n) = T\sum_{n=1}^{N} X(n, C_n)^2 \quad (4)$$

, where $N$ is the number of the observed potentials and $T$ the period of sampling.

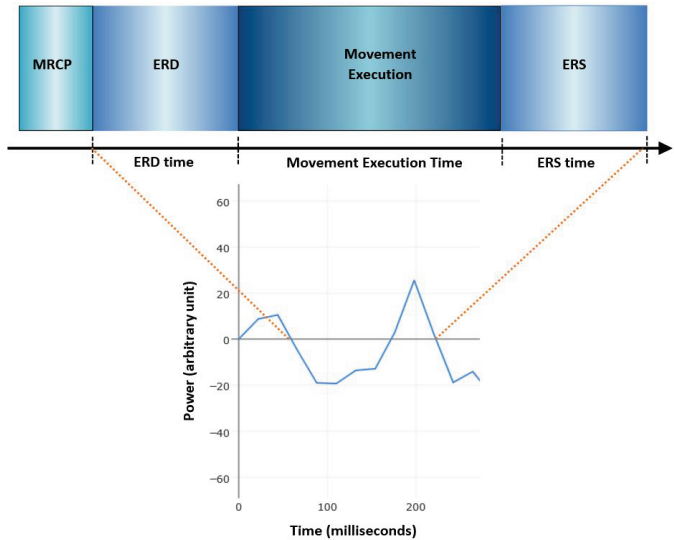

Fig. 4. The timeline of a an executed movement with ERD and ERS time along with the aditional EEG recording.

All the just described successive steps leading to the recognition of ERD/ERS patterns and the computation of their power and energy are listed in Table I.

### III. EXPERIMENTAL RESULTS OF THE PREPROCESSED TECHNIQUE

In order to evaluate the computational time and effectiveness of the proposed pre-processing technique, several experiments were conducted using prerecorded EEG datasets. These prerecorded datasets were created and contributed to PhysioNet [17] database by the developers of the BCI2000 instrumentation system [18]. Datasets includes different sessions of over 1500 of one and two minute EEG recordings that correspond to 109 volunteers. The volunteers performed different motor/imagery tasks while 64-channel EEGs were recorded. Each volunteer performed 14 experimental runs, two one-minute baseline runs (one with eyes open, one with eyes closed), and three two-minute runs of each of the tasks presented on Table II. All experiments were performed on an Intel Core i7 – 2600K at 3.70GHz machine, with 16 GB of RAM.

#### A. Offline procedure

During offline procedure, all 109 different datasets where preprocessed according to the steps described in Table I. whereas three different Blind Source Separation (BSS) algorithms were used as part of the ICA analysis. The used algorithms are: the BSS through Canonical Correlation Analysis (BSSCCA), the fast algorithm for BSS of non-Gaussian and time correlated signals (FCOMBI) and the Second Order BSS (SOBI) [19, 20]. Accordingly, for each dataset, three matrixes were produced that contain the feature vectors of Power and Energy of ERD/ERS events.

Binary Support Vector Machine (SVM) was used as the supervised learning model. In this case, all matrixes of feature vectors were combined into a new matrix that contained the values of power, and energy of ERD/ERS events, of all 109 datasets. Training data construction was based on the selection of the electrode. For each electrode $C_n$, the training data consists of the respective columns of power and energy.

TABLE I. PREPROCESSING DATA TECHNIQUE FOR REDUCING INITIAL DATA OF EEG RECORDINGS

| Preprocessing EEG data technique |
|---|
| 1. Load the data |
| 2. Select electrodes from motor cortex area |
| 3. High pass 0.5Hz ~ 90Hz to remove DC noise |
| 4. Remove EOGs and ECGs artifacts |
| 5. Band elimination @50+-5Hz to remove line noise |
| 6. Isolate epoch |
| a. Calculate total time window of movement t |
| b. Isolate the event |
| c. Reject the MRCP |
| d. Reject remaining signal |
| 7. ICA algorithm on isolated epoch |
| 8. Extract Feature Vectors for each $C_n$ electrode that contains |
| a. Power of ERD/ERS |
| b. Energy of ERD/ERS |
| c. Index that corresponds to the type of event (ERD/ERS) |
| d. Index that corresponds to the type of movement (left/right) |

TABLE II. TASKS THAT SUBJECTS PERFORMED DURING EEG RECORDINGS

| | Task Description |
|---|---|
| 1. | Open and close left or right fist |
| 2. | Imagine opening and closing left or right fist |
| 3. | open and close both fists or both feet |
| 4. | Imagine opening and closing both fists or both feet |

Two grouping variables, for the SVM algorithm, were used during the offline procedure, one between ERD/ERS events and one between left/right movements. Fig 5, illustrates the supported vectors that were produced using the offline data for electrodes $C_3$, $C_4$, $FC_3$ and $FC_z$, while Table III presents respectively the execution time of training for all the electrodes. Table IV presents the execution time of training for all the electrodes without using the described preprocess technique. Considering that the execution of the SVM training on each non preprocessed data takes several seconds, the preprocess technique reduces significantly the training time.

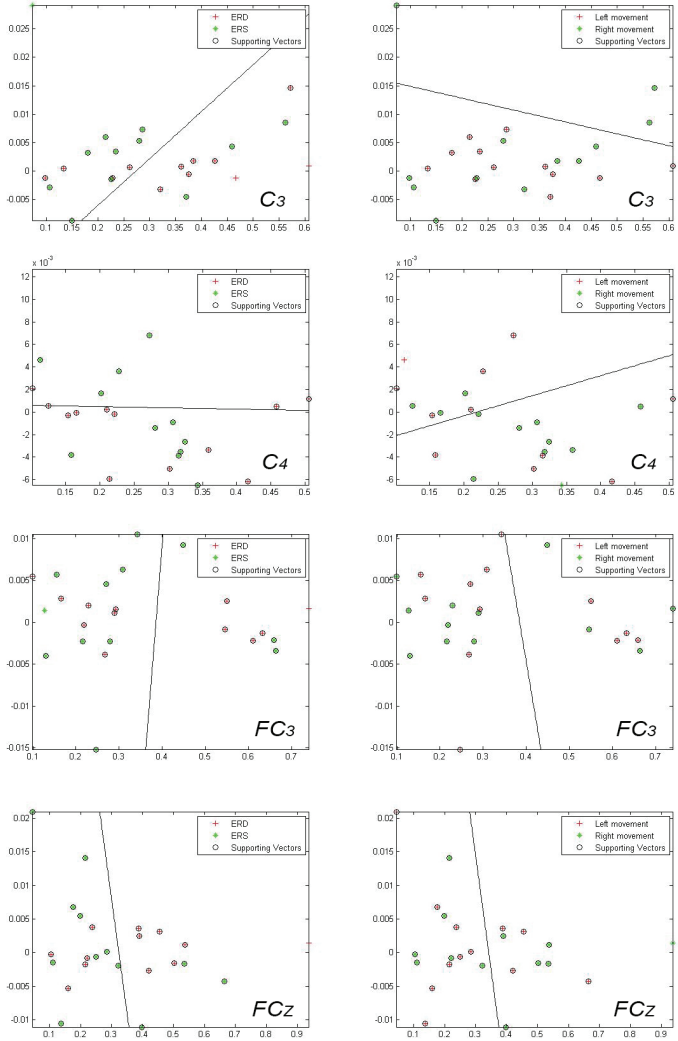

Fig. 5. The produced supported vectors of SVM during training, using the preproccess tehcnique for electrodes $C_3$, $C_4$, $FC_3$ and $FC_z$.

TABLE III. EXECUTION TIME OF SVM TRAINING USING THE PREPROCESS TECHNIQUE, IN SECONDS

| Electrode | Training Group | |
|---|---|---|
| | Left/Right, s | ERD/ERS, s |
| $FC_3$ | 0.842601 | 0.831507 |
| $FC_z$ | 0.841640 | 0.833988 |
| $FC_4$ | 0.839987 | 0.862060 |
| $C_3$ | 0.875843 | 0.874127 |
| $C_1$ | 0.857434 | 0.864836 |
| $C_z$ | 0.831123 | 0.961868 |
| $C_2$ | 0.861366 | 0.869357 |
| $C_4$ | 0.860544 | 0.861083 |

### B. Online procedure

In order to demonstrate and assess the computation time of finding the subject's imagined and executed movement type, a random dataset is selected out of the 109 datasets. The data is preprocessed the way explained in Table I and the power and energy of the first epoch are computed. This random dataset is then used as input data to the SVM supervised learning model that was used during offline procedure, in order to be classified.

Table V presents the online procedure execution time and the percentage of the falsely classified samples. Similar computation times and percentages are reported when the proposed technique is applied only to sessions related with fist, elbow or other specific motor imaginary movements. Table VI presents the classification execution time without preprocessing the data. From Tables V and VI one can easily observe that the classification time using the preprocessing technique is significantly reduced, whilst the number of erroneously classified samples is practically equal in both cases.

### IV. CONCLUSIONS

A new pre-process technique of EEG signals produced by motor imagery movements has been described. This technique focused on ERD/ERS of motor imagery phenomena, resulting to an accelerated determination of the imagery movement. Power and energy features of the processed signal are compared with the same features of classified signals from an available database. This determination is performed within the time limits imposed by the on-line control of the continuous passive rehabilitation devices. Several experiments were contacted using pre-recorded datasets demonstrating that a significant computation time reduction is achieved against a training without the described preprocess technique, in the recognition of a motor imagery movement.

TABLE IV. EXECUTION TIME OF SVM TRAINING WHITHOUT THE PREPROCESS TECHNIQUE, IN SECONDS

| Electrode | Training Group | |
|---|---|---|
| | Left/Right, s | ERD/ERS, s |
| $FC_3$ | 4.458632 | 4.234856 |
| $FC_z$ | 5.234578 | 5.365456 |
| $FC_4$ | 4.954623 | 4.942367 |
| $C_3$ | 3.589326 | 3.452356 |
| $C_1$ | 3.658456 | 3.421537 |
| $C_z$ | 3.842367 | 3.942159 |
| $C_2$ | 3.247653 | 3.478126 |
| $C_4$ | 4.635478 | 4.578236 |

TABLE V. EXECUTION TIME OF SVM CLASSIFICATION USING THE PREPROCESS TECHNIQUE IN SECONDS AND PERCENTAGE OF WRONG CLASSIFIED SAMPLES.

| Electrode | Classification Group | | | |
|---|---|---|---|---|
| | Left/Right, s | Wrong Samples, % | ERD/ERS, s | Wrong Samples, % |
| $FC_3$ | 0.124288 | 12 | 0.124459 | 11 |
| $FC_z$ | 0.125329 | 14 | 0.126518 | 13 |
| $FC_4$ | 0.128660 | 7 | 0.141144 | 5 |
| $C_3$ | 0.128067 | 8 | 0.129101 | 5 |
| $C_1$ | 0.126894 | 7 | 0.127608 | 7 |
| $C_z$ | 0.125279 | 4 | 0.125609 | 8 |
| $C_2$ | 0.127079 | 7 | 0.127313 | 12 |
| $C_4$ | 0.127945 | 9 | 0.130748 | 4 |

TABLE VI. EXECUTION TIME OF SVM CLASSIFICATION WITHOUT USING THE PREPROCESS TECHNIQUE IN SECONDS AND PERCENTAGE OF WRONG CLASSIFIED SAMPLES.

| Electrode | Classification Group | | | |
|---|---|---|---|---|
| | Left/Right, s | Wrong Samples, % | ERD/ERS, s | Wrong Samples, % |
| $FC_3$ | 5.634785 | 11 | 4.992374 | 11 |
| $FC_z$ | 4.237469 | 12 | 4.178548 | 14 |
| $FC_4$ | 4.640433 | 7 | 5.098300 | 5 |
| $C_3$ | 3.496284 | 9 | 3.994853 | 7 |
| $C_1$ | 4.195463 | 7 | 5.098300 | 5 |
| $C_z$ | 6.004934 | 4 | 6.963625 | 7 |
| $C_2$ | 5.468561 | 7 | 5.911480 | 13 |
| $C_4$ | 5.367894 | 10 | 5.732189 | 5 |

## V. FUTURE WORK

Experimental results on the pre-recorder datasets show that the described technique not only achieves a good time performance but manages to classify correct, by using the power and energy features of classified signals, real time EEG data. Still, other feature vectors such as mean power of ERD/ERS or standard deviation between the mean and observed values, can be evaluated within the described pre-processing technique, in order to perform much more accurate classification.

Furthermore, additionally datasets that correspond to several imaginary movements of human limbs, such as flexion/extension and pronation/supination of elbow, can be created and used in order to perform classification focused on a different set of imaginary human movements, within the time limits imposed by the requirements of real-time control of rehabilitation devices.